\documentclass[aps,prd,onecolumn,11pt,notitlepage,nofootinbib,showkeys]{revtex4-2}
\usepackage{amsmath,amssymb,bm}
\usepackage{bbm,slashed,microtype}
\usepackage{xcolor}
\usepackage{hyperref}
\hypersetup{colorlinks=true,linkcolor=blue,citecolor=blue,urlcolor=blue,
  pdftitle={Quaternionic Maxwell Green Function}}
\newcommand{\BB}{\mathbb{B}}
\newcommand{\CC}{\mathbb{C}}
\newcommand{\RR}{\mathbb{R}}
\newcommand{\bsig}{\vec{\sigma}}
\newcommand{\bE}{\vec{E}}
\newcommand{\bB}{\vec{B}}

\newcommand{\bk}{\vec{k}}
\newcommand{\br}{\vec{r}}
\newcommand{\brp}{\vec{r}'}

\newcommand{\bA}{\vec{A}}

\newcommand{\dd}{\mathrm d}
\newcommand{\ee}{\mathrm e}
\newcommand{\e}{\mathrm e}
\newcommand{\ii}{\mathrm i}
\newcommand{\jj}{\mathrm j}

\newcommand{\Boxop}{\Box}

\newcommand{\calA}{\mathcal A}
\newcommand{\calD}{\mathcal D}
\newcommand{\calJ}{\mathcal J}
\newcommand{\calG}{\mathcal G}
\newcommand{\calF}{\mathcal F}

\newcommand{\calR}{\mathcal R}
\newcommand{\calI}{\mathcal I}
\newcommand{\MM}{\mathbb M}

\newcommand{\PV}{\operatorname{(P.V.)}_{\rm sph}}
\begin{document}

\title{Quaternionic Maxwell Green Function}

\author{I. Heye, B. Shi and U. D. Jentschura}

\affiliation{Missouri University of Science and Technology, Rolla, Missouri, USA}

\begin{abstract}
The Pauli-matrix realization of complex quaternions combines the eight real
component equations of Maxwell theory into one first-order equation. We develop
a self-contained retarded source-to-field representation based on the 
quaternionic Maxwell Green function,
starting from Edmonds's formulation. The associated scalar Green function
generates the scalar and vector potentials together.
Applications include a moving point charge, 
the complete oscillating electric-dipole
field, and the quaternionic form of the 
Maxwell stress tensor. For oscillating fields,
an independent commuting phasor unit separates 
the harmonic phase from the
intrinsic complex structure of the electromagnetic field. Two appendices
summarize the quaternion algebra and a faithful real-matrix representation of
the bicomplex coefficient algebra. The construction is placed in the context
of earlier quaternionic fundamental solutions and recent related work.
\end{abstract}
\maketitle

\tableofcontents

\section{Introduction}

Quaternionic and complex-vector formulations offer a compact algebraic
organization of classical electrodynamics~\cite{Edmonds1978}. 
It is our goal, in the current investigations, to 
significantly advance the scope of the quaternionic formalism,
to include Green functions and the Maxwell stress tensor,
while relying on a number of previous investigations.
In fact, the rearrangement of electric
and magnetic fields into complex 
combinations is closely related to the Riemann--Silberstein vector,
whose early formulation is due to Silberstein~\cite{Silberstein1907} and whose
classical and quantum uses have been reviewed extensively
\cite{BialynickiBirula1996,BialynickiBirula2013}. Biquaternionic treatments also
have a long history, including the dissertation of Lanczos
\cite{Lanczos1919}. Edmonds \cite{Edmonds1978} presented an economical
Pauli-matrix realization in which Maxwell's equations take the form
$\calD\calF=\calJ/(\epsilon_0c)$. Here $\calD$ is a first-order differential
operator, $\calF$ packages the electric and magnetic fields, and $\calJ$
packages the charge and current densities into quaternionic combinations.

Somewhat ironically, the term ``Maxwell equations'' (plural) receives 
a new interpretation in terms of a single ``Maxwell equation'' (singular),
written as $\calD\calF=\calJ/(\epsilon_0c)$,
in the quaternionic formalism. The counting of equations is useful. 
Two of the Maxwell equations (Gauss's law and the law of the 
absence of magnetic monopoles) are scalar (one-component) 
equations, while the other equations (Faraday's law and the 
Ampere--Maxwell law) are vector equations. 
Consequently, Maxwell's four conventional vector-calculus equations contain
eight component equations. An ordinary quaternion has four real
coefficients, but the complex quaternion algebra employed here has four
complex coefficients. The real and imaginary parts of these coefficients
therefore accommodate the inhomogeneous and homogeneous Maxwell equations,
respectively. Throughout, this statement concerns coefficients in the Pauli
basis, not entrywise real and imaginary parts of a field.
The question immediately arises what to do with monochromatic
oscillating electromagnetic fields, which are customarily 
described using complex phases of the functional form 
$\exp(-\jj \, \omega \, t)$, where $\jj$ denotes an imaginary 
unit which is different from the imaginary unit $\ii$ employed in 
the combination of the electric and magnetic fields, and in 
the complex quaternionic algebra employed in the construction
of the Maxwell equation $\calD\calF=\calJ/(\epsilon_0c)$.

In the current investigations, we aim to amplify the scope 
of the quaternionic formalism by constructing the 
Green function, which, in operator language, is the inverse
$\calG = \calD^{-1}$ of the differential operator $\calD$,
and establishes the connection between the field $\calF$ and the 
sources $\calJ$ in terms of $\calF = \calG \calJ / (\epsilon_0c)$.
With retarded boundary conditions,
this gives a coordinate-space Green function supported on
the future light cone. Its convolution with a conserved current yields the
fields directly, without requiring a separate potential calculation. 
For the connection of the usual retarded source integrals
to charge conservation, we refer to Ref.~\cite{Heras2007}.

Our inverse-operator construction is related to an established body of
quaternionic analysis. Quaternionic
reformulations in inhomogeneous media were discussed by 
Kravchenko~\cite{Kravchenko2002}. Furthermore,
Grudsky, Khmelnytskaya, and Kravchenko explicitly
constructed a causal fundamental solution in a chiral medium~\cite{Grudsky2004}. 
Mohr developed a matrix Maxwell Green function in a
different, six-component representation~\cite{Mohr2010}. More recently,
Delgado and Kravchenko treated inhomogeneous time-harmonic Maxwell equations
on unbounded domains and related their solution to a dyadic Green 
function~\cite{DelgadoKravchenko2023}. A particularly close Fourier-based construction
based on biquaternionic Maxwell operators appears in the 
recent preprint of Guill\'en-Villalobos, Delgado, and Vargas Rodr\'iguez
\cite{GuillenVillalobos2026}. 
Here, we aim for a unified retarded realization of the 
quaternionic formalism, with explicit
conservation and gauge checks, complementary static and dynamical examples,
and separate intrinsic and phasor complex structures. 
Furthermore, the calculation of the quaternionic Maxwell
Green function is a significant motivation for the current
studies.

In terms of the utility of the quaternionic formalism,
we refer to recent work, specifically,
to quaternion and biquaternion representations of proper and
improper transformations \cite{KatrusiakLe2024}, and matrix equations over
commutative quaternion rings \cite{Chen2025,Chang2026}. The latter terminology
refers to a commutative four-real-dimensional algebra and must not be confused
with the noncommutative quaternions. 

The current study is organized as follows. 
Section~\ref{sec2} introduces the quaternionic Maxwell equation, its momentum-
and coordinate-space inverses, and the relation to potentials and gauge
freedom. Section~\ref{sec3} treats several applications, 
including moving point charges, 
electric-dipole radiation,
and the Maxwell stress tensor. Conclusions are given in Section~\ref{sec4}.
Appendix~\ref{app:quaternions} fixes the quaternion conventions, and
Appendix~\ref{app:bicomplex} constructs the real-matrix representation of two
commuting complex structures (bicomplex structures, which are essential
for our treatment of the dipole radiation).

\section{Quaternionic Maxwell Equation and Green Function}
\label{sec2}

\subsection{Quaternionic Equation and Conventions}
\label{sec21}

We work in vacuum in SI units, with $c^{-2}=\epsilon_0\mu_0$, and use the
Pauli matrices $\sigma_0=\mathbbm{1}_{2 \times 2}$ and 
$\sigma_a$, $a=1,2,3$. Our starting point is
Edmonds's equation \cite{Edmonds1978}, with his conventional overall factor
$\ii\hbar$ divided out:
\begin{equation}
 \calD\calF=\frac{1}{\epsilon_0c}\calJ.
 \label{eq:Edmonds}
\end{equation}
The operator $\calD$, the field $\calF$, and 
the source $\calJ$ are
\begin{equation}
 \label{eq:DFJdef}\\
 \calD =\frac{\sigma_0}{c}\partial_t-\bsig\cdot\vec{\nabla}, \qquad
 \calF =-\bsig\cdot(\bE-\ii c\bB), \qquad
 \calJ =c\rho\sigma_0+\bsig\cdot\vec{J}.
\end{equation}
Thus, $P=\ii\hbar\calD$ and $\epsilon'=\ii\hbar/(\epsilon_0c)$ in Edmonds's
notation~\cite{Edmonds1978}.
We have rewritten the Maxwell equation
so that the quantum-mechanical constant $\hbar$
is eliminated from the classical field equation. 
The physical time-domain fields and sources in
Eq.~\eqref{eq:DFJdef} are real rather than complex
{\em with respect to the imaginary unit $\ii$,
but can be complex with respect to an 
additional imaginary unit $\jj$ which will be 
introduced below.}
The Pauli identity given in Eq.~\eqref{eq:PauliIdentity}
of Appendix~\ref{app:quaternions} gives the expansion
\begin{align}
 \calD\calF={}&
 (\vec{\nabla}\cdot\bE-\ii c\vec{\nabla}\cdot\bB)\sigma_0
 \nonumber\\
 &+\bsig\cdot\left[
 c\vec{\nabla}\times\bB-\frac1c\partial_t\bE
 +\ii(\vec{\nabla}\times\bE+\partial_t\bB)\right]
 =\mathcal R+\ii\mathcal I,
 \label{eq:DFsplit}
\end{align}
where the real part $\calR$ and the imaginary part
$\calI$ (taken with respect to the imaginary unit $\ii$) 
are given by
\begin{subequations}
\begin{align}
 \mathcal R&=(\vec{\nabla}\cdot\bE)\sigma_0
   +c\bsig\cdot\left(\vec{\nabla}\times\bB-c^{-2}\partial_t\bE\right),
 \label{eq:realpart}\\
 \mathcal I&=-c(\vec{\nabla}\cdot\bB)\sigma_0
   +\bsig\cdot(\vec{\nabla}\times\bE+\partial_t\bB).
 \label{eq:imagpart}
\end{align}
\end{subequations}
Equating these coefficients with the source $\calJ$,
we obtain $\mathcal R=\calJ/(\epsilon_0c)$ and $\mathcal I=0$, or explicitly
\begin{subequations}
\begin{align}
 \vec{\nabla}\cdot\bE&=\frac{\rho}{\epsilon_0},&
 \vec{\nabla}\times\bB-c^{-2}\partial_t\bE&=\mu_0\vec{J},
 \label{eq:inhomogeneous}\\
 \vec{\nabla}\cdot\bB&=0,&
 \vec{\nabla}\times\bE+\partial_t\bB&=0.
 \label{eq:homogeneous}
\end{align}
\end{subequations}
All eight Maxwell equations, in their 
well-known form, are therefore present. 
For the following 
considerations, we need to 
carefully distinguish 
between ordinary complex conjugation, Hermitian conjugation, and quaternionic
conjugation. For a Pauli expansion, the last is
\begin{equation}
 Q=q_0 \, \sigma_0+\vec{q}\cdot\bsig,
\qquad 
Q^\ddagger=q_0 \, \sigma_0-\vec{q}\cdot\bsig.
\label{eq:qconj}
\end{equation}
It leaves central scalar coefficients unchanged, including $\ii$, and
reverses products: $(QR)^\ddagger=R^\ddagger Q^\ddagger$. In particular,
$\calF^\ddagger=-\calF$. This convention is consistent with the ordinary
quaternionic algebra described in Appendix~\ref{app:quaternions}.
The transformation properties
under Lorentz transformations are discussed explicitly by Edmonds
\cite{Edmonds1978}; the connection with complex electromagnetic vectors and
Dirac algebra is also discussed in Ref.~\cite{Hacyan2019}.

\subsection{Momentum-Space Representation}
\label{sec22}

Fourier analysis introduces phase information in addition to the intrinsic
$\ii$ already used in the Pauli algebra. To keep these roles separate, we use
a central commuting unit $\jj$ for all Fourier phases, with
$\jj^2=-1$ and $\ii \, \jj=\jj \, \ii$. The algebraic details are given in
Sec.~\ref{sec3B} and Appendix~\ref{app:bicomplex}. The Fourier
convention is
\begin{equation}
X(\br,t)=\int\frac{\dd\omega\,\dd^3k}{(2\pi)^4}
\exp\left[ \jj \, \left( \bk\cdot\br-\omega t \right) \right] \,
\widetilde X(\omega,\bk).
\label{eq:Fourier}
\end{equation}
Consequently, $\partial_t\mapsto-\jj \, \omega$, 
$\vec{\nabla}\mapsto\jj\bk$, and
\begin{equation}
 \calD(\omega,\bk)=-\jj\left(\frac{\omega}{c}\sigma_0+\bsig\cdot\bk\right),
 \qquad
 \calD^\ddagger(\omega,\bk)=-\jj\left(\frac{\omega}{c}\sigma_0-\bsig\cdot\bk\right).
 \label{eq:Pmomentum}
\end{equation}
Since $(\bsig\cdot\bk)^2=k^2\sigma_0$, with $k=|\bk|$, one has
\begin{equation}
 \calD\calD^\ddagger=
 \left(k^2-\frac{\omega^2}{c^2}\right)\sigma_0,
 \label{eq:factor_mom}
\end{equation}
and, away from the light cone, one can invert the operator $\calD$
easily, obtaining an expression which is structurally 
similar to the fermionic Feynman propagator of quantum
electrodynamics,
\begin{equation}
 \calD^{-1}(\omega,\bk)=\jj\,
 \frac{(\omega/c)\sigma_0-\bsig\cdot\bk}{(\omega/c)^2-k^2}.
 \label{eq:Pinverse}
\end{equation}
In fact, this expression can be identified as
the two-component massless counterpart of the algebraic inversion
of a Dirac operator, $(\slashed p-m)^{-1}=(\slashed p+m)/(p^2-m^2)$, before
specifying a pole prescription; see Ref.~\cite{JentschuraAdkins2022}. The
analogy concerns the operator factorization and, of course,
does not imply a change of the spin of the
classical electromagnetic field toward that of the 
light fermions.

The retarded inverse is the boundary value from the upper half of the
$\CC_{\jj}$ frequency plane.
Denoting by $\eta$ a positive infinitesimal quantity, one obtains
for the Fourier transform of the Green function, 
which we denote by a tilde for absolute clarity,
\begin{equation}
 \widetilde{\calG}_R(\omega,\bk)
 =\lim_{\eta\downarrow0}\jj\,
 \frac{[(\omega+\jj \, \eta)/c]\sigma_0-\bsig\cdot\bk}
 {[(\omega+\jj \, \eta)/c]^2-k^2}.
 \label{eq:GPmom}
\end{equation}
Both poles, $\omega=\pm ck-\jj \, \eta$, are below the real axis,
as is required for a retarded Green function. The field
of a conserved source with no added incoming or free field is
therefore
\begin{equation}
 \widetilde{\calF} = \frac{1}{\epsilon_0c} \,
 \widetilde{\calG}_R \, \widetilde{\calJ}.
 \label{eq:solution}
\end{equation}
The scalar part of the numerator of the 
expression $\widetilde{\calG}_R\widetilde{\calJ}$ is 
proportional to 
\begin{equation}
 \omega\widetilde\rho-\bk\cdot\widetilde{\vec{J}}=0,
 \label{eq:fourier-continuity}
\end{equation}
which is the Fourier transform of charge conservation. Thus the inverse
produces a pure Pauli-vector field, as required by Eq.~\eqref{eq:DFJdef}.

\subsection{Coordinate-Space Green Function}
\label{sec33}

In coordinate space, the 
quaternion conjugate differential operator $\calD^\ddagger$ is 
\begin{align}
 \calD^\ddagger=& \; \frac{\sigma_0}{c}\partial_t+\bsig\cdot\vec{\nabla},
 \label{eq:Dddagger}
\\
 \calD\calD^\ddagger=& \; \calD^\ddagger\calD=\Boxop\sigma_0,
 \qquad \Boxop=c^{-2}\partial_t^2-\vec{\nabla}^2.
 \label{eq:factorcoord}
\end{align}
It follows that the retarded first-order inverse factors as
\begin{equation}
 \calG_R=\calD_R^{-1}=\calD^\ddagger G_R,
 \qquad G_R=\Boxop_R^{-1}.
 \label{eq:Pinversecoord}
\end{equation}
Here, an inverse operator is identified with its convolution kernel. 
One obtains
\begin{equation}
 G_R(\vec r - \vec r' , t - t') =
\frac{\delta(t - t' - | \vec r - \vec r'|/c)}{4\pi \, |\vec r - \vec r'|},
 \qquad
 \Boxop G_R=\delta(t - t')\delta^{(3)}(\vec r - \vec r') \,.
 \label{eq:scalarGR}
\end{equation}
A contour derivation is given in Sec.~4.1.3 of
Ref.~\cite{Jentschura2017}, notably Eqs.~(4.27)--(4.35). The Green function
in that reference includes an additional factor $1/\epsilon_0$; our kernel
is $\epsilon_0$ times that convention.  Indeed,
one finds that 
\begin{equation}
 \calD\calG_R=(\Boxop G_R)\sigma_0
 =\delta(t - t') \, \delta^{(3)}(\vec r - \vec r') \, \sigma_0.
 \label{eq:green-check}
\end{equation}
The direct retarded solution is
\begin{equation}
 \calF(\br,t)=\frac{1}{\epsilon_0c}
 \int\dd t'\,\dd^3r'\,
 \calG_R(\br-\brp,t-t')\calJ(\brp,t').
 \label{eq:field-convolution}
\end{equation}

In a static case, the quaternionic formalism becomes 
particularly instructive; the 
static fields can be found without introducing potentials. Setting
$\partial_t=0$ gives
\begin{equation}
 \calD_s=-\bsig\cdot\vec{\nabla}.
 \label{eq:Ps}
\end{equation}
With the Fourier convention in Eq.~\eqref{eq:Fourier}, its 
Fourier transform and its inverse are
\begin{equation}
 \calD_s(\bk)=-\jj \, \bsig\cdot\bk,
 \qquad\calD_s^{-1}(\bk) = \jj \, \frac{\bsig\cdot\bk}{k^2}.
 \label{eq:Psinv}
\end{equation}
The coordinate-space kernel is therefore
\begin{equation}
\calG_s(\vec r - \vec r') =
\bsig\cdot\vec{\nabla}_{\vec r} \, 
\frac1{4\pi | \vec r - \vec r'|}
\qquad
\calD_s\calG_s(\vec r - \vec r') =\delta^{(3)}(\vec r - \vec r') \, \sigma_0.
\label{eq:staticfirstorder}
\end{equation}
The quaternionic field is
%
\begin{align}
 \calF(\br)= & \; \frac1{\epsilon_0c}\int\dd^3r'\,
 \calG_s(\br-\brp)\calJ(\brp)
\nonumber\\
=& -\frac{1}{4\pi\epsilon_0c} \,
\int\dd^3r' \frac{(\vec r - \vec r') \cdot\vec{J}(\vec r')}%
{| \vec r - \vec r'|^3}\sigma_0
-\bsig\cdot\frac1{4\pi\epsilon_0}\int\dd^3r'
\frac{(\vec r - \vec r') \, \rho(\brp)}{ | \vec r - \vec r'|^{3} }
\nonumber\\
&+\ii c\,\bsig\cdot\frac{\mu_0}{4\pi}\int\dd^3r'
\frac{\vec{J}(\vec r') \times (\vec r - \vec r') }{ | \vec r - \vec r' |^3}.
\end{align}
The scalar part vanishes for localized conserved currents because
\begin{equation}
 \int\dd^3r'\frac{ ( \vec r - \vec r') \cdot\vec{J}(\vec r')}%
{ | \vec r - \vec r' |^3 } 
 =\int\dd^3r'\,\vec{J}(\vec r') \cdot
\vec{\nabla}'\frac{1}{ | \vec r - \vec r' |} 
 =-\int\dd^3r' \frac{1}{ | \vec r - \vec r' |} \, 
\vec{\nabla}'\cdot\vec{J}(\vec r') = 0 \,.
 \label{eq:static-scalar-zero}
\end{equation}
The remaining coefficients give 
\begin{align}
 \bE(\br)&=\frac1{4\pi\epsilon_0}\int\dd^3r'
 \rho(\brp)\frac{\br-\brp}{|\br-\brp|^3},
 \label{eq:Coulombfromstatic}\\
 \bB(\br)&=\frac{\mu_0}{4\pi}\int\dd^3r'
 \frac{\vec{J}(\brp)\times(\br-\brp)}{|\br-\brp|^3} \, ,
 \label{eq:BiotSavartfromstatic}
\end{align}
and they 
determine Coulomb's law and the Biot--Savart law in a single calculation.

\subsection{Scalar and Vector Potentials}
\label{sec24}

Up to now, we have directly obtained the electromagnetic 
fields from the sources. This raises the question of how
to incorporate scalar and vector potentials into the 
quaternionic formalism. We define the quaternionic four-potential 
by
\begin{equation}
 \calA=\frac{\Phi}{c}\sigma_0+\bsig\cdot\bA.
 \label{eq:Aquat}
\end{equation}
Its quaternionic derivative has the expansion
\begin{equation}
 \calD^\ddagger\calA=
 \left(\vec{\nabla}\cdot\bA+c^{-2}\partial_t\Phi\right)\sigma_0
 +\bsig\cdot\left[c^{-1}(\partial_t\bA+\vec{\nabla}\Phi)
 +\ii\vec{\nabla}\times\bA\right].
 \label{eq:DdagA}
\end{equation}
With $\bE=-\vec{\nabla}\Phi-\partial_t\bA$ and $\bB=\vec{\nabla}\times\bA$, this is
\begin{equation}
\calD^\ddagger\calA=\Lambda\sigma_0+\frac{\calF}{c},
\qquad\Lambda = \vec{\nabla}\cdot\bA + c^{-2} \, \partial_t \Phi \, .
\label{eq:DdagAcompact}
\end{equation}
Edmonds already includes a potential--field relation and the scalar
quaternionic projection in his Eq.~(4) \cite{Edmonds1978}. 
We consider the following scalar-kernel convolution,
which is reminiscent of the way in which the
field equations are solved in Lorenz gauge~\cite{Jentschura2017}, 
\begin{equation}
\calA_R(\vec{r},t)=\mu_0 \int \dd t' \int \dd^3 r' \,
G_R(\vec r - \vec r', t - t') \, \calJ(\vec r', t') \,.
 \label{eq:Aretconv}
\end{equation}
The scalar and vector coefficients satisfy
\begin{subequations}
\label{eq:potential-components}
\begin{align}
\label{eq:potential-component scalar} 
\Phi_R(\vec r, t) =& \; 
\frac{1}{\epsilon_0} \int\dd t'\,\dd^3r'\,G_R(\vec{r}-\vec{r}',t-t')\rho(\vec{r}',t') \,,
\\
\bA_R(\vec r, t) =& \; 
\mu_0\int\dd t'\,\dd^3r'\,G_R(\vec{r}-\vec{r}',t-t')\vec{J}(\vec{r}',t'),
\label{eq:potential-component vector}
\end{align}
\end{subequations}
 
{\rm If} we assume that the quaternionic potential 
$\calA_R(\vec{r},t)$ is given by Eq.~\eqref{eq:Aretconv},
then it is easy to show that the function
$\Lambda_R = \vec{\nabla}\cdot\bA + c^{-2} \, \partial_t \Phi$ 
takes the form
\begin{equation}
 \Lambda_R=\mu_0\int\dd t'\,\dd^3r'\,
 G_R(\vec{r}-\vec{r}',t-t')\left[\partial_{t'}\rho(\vec{r}',t')+\vec{\nabla}'\cdot\vec{J}(\vec{r}',t')\right]=0.
 \label{lorentz3}
\end{equation}
Thus, the retarded potentials 
{\em calculated according to Eq.~\eqref{eq:Aretconv}}
are naturally in Lorenz gauge, without a
separate gauge adjustment. The time integration gives the standard forms
\begin{align}
 \Phi_R(\br,t)&=\frac{1}{4\pi\epsilon_0}\int\dd^3r'
 \frac{\rho(\brp,t-|\br-\brp|/c)}{|\br-\brp|},
 \label{eq:retphi}\\
 \bA_R(\br,t)&=\frac{\mu_0}{4\pi}\int\dd^3r'
 \frac{\vec{J}(\brp,t-|\br-\brp|/c)}{|\br-\brp|}.
 \label{eq:retA}
\end{align}
This is compatible with the
conventional potential formulation
outlined in Refs.~\cite{Jackson1999,Jentschura2017}.
In Lorenz gauge,
\begin{equation}
 \calF=c\calD^\ddagger\calA_R
 =\frac{1}{\epsilon_0c}(\calD^\ddagger G_R)*\calJ
 =\frac{1}{\epsilon_0c}\calG_R*\calJ \,,
 \label{eq:connection}
\end{equation}
where the asterisk ($*$) denotes convolution.
The field obtained from this expression is gauge invariant.
For arbitrary gauge, the relation between the 
quaternionic field $\calF$ and the potential $\calA$
is not as simple, but we can write
\begin{equation}
 \calF=\frac c2\left[\calD^\ddagger\calA
 -(\calD^\ddagger\calA)^\ddagger\right].
 \label{eq:gaugeindependentprojection}
\end{equation}
Independently of gauge, the potential definition gives
$\vec{\nabla}\cdot\bB=0$ and
$\vec{\nabla}\times\bE+\partial_t\bB=0$ identically.

\section{Applications of the Quaternionic Formalism}
\label{sec3}

\subsection{Application I: Moving Point Charge}
\label{sec3A}

Let a charge $q$ move along a differentiable subluminal trajectory $\vec{R}(t)$.
The charge and current densities are
\begin{equation}
\label{eq:movingrhoj}
 \rho(\br,t) = q\delta^{(3)}[\br-\vec{R}(t)], \qquad
 \vec{J}(\br,t) =q\dot{\vec{R}}(t)\delta^{(3)}[\br-\vec{R}(t)] \,,
\end{equation}
so that the quaternionic source is
\begin{equation}
 \calJ(\br,t)=q\left[c\sigma_0+\bsig\cdot\dot{\vec{R}}(t)\right]
 \delta^{(3)}[\br-\vec{R}(t)].
 \label{eq:movingJ}
\end{equation}
The continuity equation follows distributionally by differentiating the
moving delta function. We use the 
Lorenz-gauge quaternionic potential 
formulation based on Eq.~\eqref{eq:Aretconv} and
first perform the spatial integral:
\begin{equation}
 \calA(\br,t)=\frac{\mu_0q}{4\pi} \, 
 \int\dd t' \,
 \frac{c\sigma_0+\bsig\cdot\dot{\vec{R}}(t')}{s(t')} \,
 \delta\!\left(t-t'-\frac{s(t')}{c}\right),
 \label{eq:Apointpre}
\end{equation}
where $\vec{s}(t')=\br-\vec{R}(t')$ and $s(t')=|\vec{s}(t')|$. The retarded
root and associated geometrical quantities are
\begin{subequations}
\begin{align}
 t-t_r= & \; \frac{|\br-\vec{R}(t_r)|}{c},
 \qquad\vec{\rho}=\br-\vec{R}(t_r),\qquad\rho=|\vec{\rho}|,
 \label{eq:retardedtime}
\\
 \vec{n}=& \; \frac{\vec{\rho}}{\rho},\qquad
 \vec{\beta}=\frac{\dot{\vec{R}}(t_r)}c,\qquad
 \varkappa=1-\vec{n}\cdot\vec{\beta}>0.
 \label{eq:kappa}
\end{align}
\end{subequations}
For $f(t')=t-t'-s(t')/c$, one finds $f'(t_r)=-\varkappa$. Provided the
observation event has a retarded intersection with the particle's trajectory,
monotonicity for a subluminal source makes that root unique, and
it follows that $\delta[f(t')]=\delta(t'-t_r)/\varkappa$
and that the quaternionic potential is given by
\begin{equation}
 \calA(\br,t)=\frac{\mu_0q}{4\pi}
 \left[\frac{c\sigma_0+\bsig\cdot\dot{\vec{R}}}{\varkappa\rho}\right]_{\rm ret}.
 \label{eq:Apoint}
\end{equation}
The components are the Lorenz-gauge Li\'enard--Wiechert potentials,
\begin{equation}
 \Phi(\br,t)=\frac{q}{4\pi\epsilon_0}\left[\frac1{\varkappa\rho}\right]_{\rm ret},
 \qquad
 \bA(\br,t)=\frac{\mu_0q}{4\pi}\left[\frac{\dot{\vec{R}}}{\varkappa\rho}\right]_{\rm ret}.
 \label{eq:LWpotentials}
\end{equation}
Their characteristic denominator is therefore the Jacobian of the retarded
time constraint, not an additional ansatz. Standard discussions are given
in Refs.~\cite{Jackson1999,Jentschura2017}.
Differentiating the potential, we use the 
relations $\partial_t t_r=\varkappa^{-1}$ and 
$\vec{\nabla} t_r=-\frac{\vec{n}}{c\varkappa}$,
apply $c\calD^\ddagger$, and obtain (for points off the world line),
\begin{align}
 \bE(\br,t)=& \; \frac{q}{4\pi\epsilon_0}
 \left[
 \frac{(1-\beta^2)(\vec{n}-\vec{\beta})}{\varkappa^3\rho^2}
 +\frac{\vec{n}\times[(\vec{n}-\vec{\beta})\times\dot{\vec{\beta}}]}
 {c\varkappa^3\rho}
 \right]_{\rm ret},
 \label{eq:LWE} \\
 \bB(\br,t)=& \; \frac1c\vec{n}\times\bE(\br,t).
 \label{eq:LWB}
\end{align}
Here $\dot{\vec{\beta}}=\dd\vec{\beta}/\dd t_r$, and $\vec{n}$ in
Eq.~\eqref{eq:LWB} is the retarded direction defined in
Eq.~\eqref{eq:kappa}. 
A stationary charge gives the
Coulomb field, while constant velocity removes the acceleration term.
The same result follows directly from Eq.~\eqref{eq:field-convolution};
the potential route makes the retarded-time Jacobian particularly explicit.

\subsection{Application II: Electric--Dipole Radiation}
\label{sec3B}

{\em Bicomplex Extension of the Formalism.---}The 
calculation of the oscillating
electric dipole illustrates a distinction between two
complex structures. In real time-domain electrodynamics, the central unit
$\ii$ in $\calF=-\bsig\cdot\bE+\ii c\bsig\cdot\bB$ packages two real
fields and appears in the Pauli multiplication rule. Harmonic amplitudes
introduce a second complex structure that records temporal phase. We retain
both structures independently rather than identifying them implicitly.
The coefficient algebra is bicomplex,
\begin{equation}
 \BB=\RR[\ii,\jj]\simeq\CC_{\ii}\otimes_{\RR}\CC_{\jj},
 \qquad \ii^2=\jj^2=-1,\qquad
   \ii \, \jj=\jj \, \ii \,.
 \label{eq:bicomplex_algebra}
\end{equation}
This is the bicomplex algebra, not Hamilton's quaternion algebra;
we refer to Appendix~\ref{app:bicomplex} for some explanatory remarks.
General treatments are given in
Ref.~\cite{Ronn2001}. Related commutative quaternion matrix algebras occur in
Refs.~\cite{Chen2025,Chang2026}. 
For $\omega>0$, a monochromatic field is written as
\begin{align}
 \bE(\br,t)&=\operatorname{Re}_{\jj}
   [\vec{E}_0(\br)\e^{-\jj\omega t}],&
 \bB(\br,t)&=\operatorname{Re}_{\jj}
   [\vec{B}_0(\br)\e^{-\jj\omega t}],
 \label{eq:phasor_fields_j}
\end{align}
where $\vec{E}_0$ and $\vec{B}_0$ have three $\jj$-complex
entries, {\em i.e.}, $\vec{E}_0,\vec{B}_0\in\CC_{\jj}^3$. 
Sources are represented in
the same way. The generalized field and current amplitudes are
\begin{equation}
 \calF_0=-\bsig\cdot(\vec{E}_0 - \ii c\vec{B}_0),
 \qquad
 \calJ_0=c \, \rho_0 \, \sigma_0+\bsig\cdot\vec{J}_0.
 \label{eq:F0_bicomplex}
\end{equation}
The amplitudes themselves do not contain the factor $\e^{-\jj\omega t}$.
Phasor conjugation, denoted by ${}^{*_{\jj}}$, reverses $\jj$ but leaves
$\ii$ and every Pauli matrix unchanged:
\begin{equation}
 \operatorname{Re}_{\jj}Q=\tfrac12(Q+Q^{*_{\jj}}).
 \label{eq:j_conjugation}
\end{equation}
Quaternionic conjugation remains
$(q_0\sigma_0+\bsig\cdot\vec{q})^\ddagger
=q_0\sigma_0-\bsig\cdot\vec{q}$ and leaves both central units fixed.
A further operation is useful for separating the electric and magnetic
sectors. One defines
the \emph{coefficientwise} intrinsic conjugation
\begin{equation}
 Q^{*_{\ii}} \equiv C_{\ii}(Q)=\sum_{\mu=0}^3q_\mu^{*_{\ii}}\sigma_\mu,
 \qquad Q=\sum_{\mu=0}^3q_\mu\sigma_\mu,
 \label{eq:intrinsic-conjugation}
\end{equation}
where ${}^{*_{\ii}}$ conjugates $\ii$ while fixing $\jj$. Because the Pauli
basis is held fixed, this is not entrywise complex conjugation of the
$2\times2$ matrix. It is the intrinsic Hermitian adjoint (with $\jj$ fixed),
and reverses products: $C_{\ii}(QR)=C_{\ii}(R)C_{\ii}(Q)$.
In contrast, phasor conjugation 
($\jj$ conjugation) preserves the order of products. The fields
can be recovered unambiguously from
\begin{equation}
 \bsig\cdot\vec{E}_0=-\frac{\calF_0+C_{\ii}(\calF_0)}2,
 \qquad
 c\bsig\cdot\vec{B}_0=\frac{\calF_0-C_{\ii}(\calF_0)}{2\ii}.
 \label{eq:extract-fields}
\end{equation}

Replacing $\partial_t$ by $-\jj\omega$ in the normalized Maxwell operator
preserves the coefficient decomposition
into $\ii$-real and $\ii$-imaginary parts,
\begin{equation}
 \calD_\omega\calF_0=\mathcal R_\omega+\ii\mathcal I_\omega,
 \qquad \calD_\omega=-\jj k\sigma_0-\bsig\cdot\vec{\nabla},
 \qquad k=\omega/c,
 \label{eq:RplusI}
\end{equation}
where
\begin{align}
 \mathcal R_\omega&=(\vec{\nabla}\cdot\vec{E}_0)\sigma_0
 +\bsig\cdot\left(c\vec{\nabla}\times\vec{B}_0
 +\frac{\jj\omega}{c}\vec{E}_0\right),
 \label{eq:Romega}\\
 \mathcal I_\omega&=-c(\vec{\nabla}\cdot\vec{B}_0)\sigma_0
 +\bsig\cdot(\vec{\nabla}\times\vec{E}_0-\jj\omega\vec{B}_0).
 \label{eq:Iomega}
\end{align}
These quantities have $\CC_{\jj}$ coefficients in the Pauli basis. Since
$\calJ_0$ has no intrinsic $\ii$-odd coefficient, the field equation gives
\begin{align}
 \vec{\nabla}\cdot\vec{E}_0&=\rho_0/\epsilon_0,&
 \vec{\nabla}\times\vec{B}_0+\frac{\jj\omega}{c^2}\vec{E}_0&=\mu_0\vec{J}_0,
 \label{eq:inhomogeneous_phasor_j}\\
 \vec{\nabla}\cdot\vec{B}_0&=0,&
 \vec{\nabla}\times\vec{E}_0&=\jj\omega\vec{B}_0.
 \label{eq:homogeneous_phasor_j}
\end{align}
Thus, the homogeneous and inhomogeneous equations remain distinct even
when the source and field amplitudes carry arbitrary harmonic phases.

{\em Electric--Dipole Source and Potentials.---}We 
first treat the ideal electric point dipole, for which the result is exact
outside the origin. For a source of characteristic size $a$, this is its
leading electric-dipole contribution under the conditions $ka\ll1$ and
$r\gg a$, with no initial assumption that $kr\gg1$. Higher electric and
magnetic multipoles are excluded from this approximation. The conventional
potential and multipole treatments are discussed in
Refs.~\cite{Jackson1999,Jentschura2017}.
For a localized monochromatic source, the continuity equation reads
in the mixed frequency-coordinate representation,
\begin{equation}
 -\mathrm j \, \omega \, \rho_0(\vec r) +
 \vec{\nabla}\cdot\vec{J}_0(\vec r) = 0 \,.
 \label{eq:edm-continuity}
\end{equation}
The electric-dipole amplitude is
\begin{equation}
 \vec{p}_0 = \int \mathrm d^3r\,
 \vec{r}\, \rho_0(\vec{r}).
\end{equation}
Integration by parts, with vanishing surface terms, gives
\begin{equation}
\int \mathrm d^3r\,\vec{J}_0 =
-\int \mathrm d^3r\, \vec{r}\, ( \vec{\nabla}\cdot\vec{J}_0 )
= -\mathrm j\omega\vec{p}_0 \, .  
\label{eq:edm-integrated-current}
\end{equation}
An ideal electric point dipole at the origin is represented by
\begin{equation}
\rho_0(\vec{r}) = -\vec{p}_0\cdot\vec{\nabla}\delta^{(3)}(\vec{r}),
\qquad
\vec{J}_0(\vec{r}) = -\mathrm j\omega\vec{p}_0\,\delta^{(3)}(\vec{r}).
\label{eq:edm-point-source}
\end{equation}
These distributions satisfy Eq.~\eqref{eq:edm-continuity} exactly.
The outgoing Green function gives the complete ideal-dipole potentials.
For the convention $\ee^{-\mathrm j\omega t}$, the outgoing Helmholtz
Green function is
\begin{equation}
G_k(r) = \frac{e^{\mathrm jkr}}{4\pi r},
\qquad
(\vec{\nabla}^2+ k^2) \, G_k(\vec r) = -\delta^{(3)}(\vec{r}) \, .
\label{eq:edm-green}
\end{equation}
Using Eq.~\eqref{eq:potential-components}, the retarded potential amplitudes are
$\Phi_0(\vec{r}) = \frac{1}{\epsilon_0}
 \int \mathrm d^3r'\, G_k(|\vec{r}-\vec{r}'|) \rho_0(\vec{r}')$,
and $\vec{A}_0(\vec{r}) =
\mu_0 \int \mathrm d^3r'\, G_k(|\vec{r}-\vec{r}'|) \vec{J}_0(\vec{r}')$.
For the source in Eq.~\eqref{eq:edm-point-source}, writing
\begin{equation}
 \vec{n}=\frac{\vec{r}}{r},
 \qquad
 \vec{\nabla} G_k(\vec r) = 
 \left( \jj \, k-\frac1r \right)\, \vec{n} \, G_k(\vec r) \, ,
 \label{eq:edm-green-gradient}
\end{equation}
we obtain the quaternionic potential 
$\calA_0(\vec r)$ from 
\begin{subequations}
\begin{align}
\Phi_0(\vec{r}) =& \;
-\frac{1}{\epsilon_0} \vec{p}_0\cdot\vec{\nabla} G_k(r)
= \frac{e^{\mathrm jkr}}{4\pi\epsilon_0}
(\vec{n}\cdot\vec{p}_0) \left(
\frac1{r^2}-\frac{\mathrm j \, k}{r} \right),
\label{eq:edm-scalar-potential}
\\
\vec{A}_0(\vec{r}) 
=& \; -\mathrm j\mu_0\omega\,\vec{p}_0G_k(r)
= -\frac{\mathrm j\mu_0\omega}{4\pi r}
\vec{p}_0 e^{\mathrm jkr} \,, 
\label{eq:edm-vector-potential}
\\
\calA_0(\vec r) =& \; 
\frac{\Phi_0(\vec r)}{c}\sigma_0+\bsig\cdot\bA_0(\vec r) \,.
\end{align}
\end{subequations}
The scalar potential contains both a $1/r^2$ contribution and a
$1/r$ contribution.  The former must not be discarded before
checking the Lorenz condition or calculating the full electric field.

{\em From the Quaternionic Potential to the Fields.---}We 
use the quaternionic representation
\begin{equation}
\mathcal A_0(\vec r) =
\mu_0 \, c \, 
\left[ \left( \frac1r-\mathrm jk \right) (\vec{n}\cdot\vec{p}_0)\sigma_0
- \mathrm j \, k \, \vec{p}_0\cdot\vec{\sigma} \right] \, G_k(\vec r) \,,
\label{eq:edm-quaternion-potential-explicit}
\end{equation}
and recall that we calculated the potentials in the 
Lorenz-gauge formalism,
\begin{equation}
\vec{\nabla}\cdot\vec{A}_0 -
\frac{\mathrm j\omega}{c^2}\Phi_0 = 0 \,.
\label{eq:edm-lorenz}
\end{equation}
The frequency-domain conjugate differential operator is
$\calD_\omega^\ddagger = -\mathrm jk\,\sigma_0
 + \vec{\sigma}\cdot\vec{\nabla}$.
Its action on the potential gives
\begin{equation}
c\calD_\omega^\ddagger\mathcal A_0
= c\left( \vec{\nabla}\cdot\vec{A}_0
 - \frac{\mathrm j\omega}{c^2}\Phi_0
 \right) \, \sigma_0
 + \vec{\sigma}\cdot \left[
 \vec{\nabla}\Phi_0 -
 \mathrm j\omega\vec{A}_0 +
 \mathrm ic\,\vec{\nabla}\times\vec{A}_0
 \right] \, .
 \label{eq:edm-operator-expanded}
\end{equation}
The scalar component vanishes by Eq.~\eqref{eq:edm-lorenz}.  Therefore
\begin{equation}
 \calF_0 = c\calD_\omega^\ddagger\mathcal A_0
 = -\vec{E}_0\cdot\vec{\sigma} +
 \mathrm ic\,\vec{B}_0\cdot\vec{\sigma},
 \label{eq:edm-field-from-potential}
\end{equation}
where
\begin{equation}
 \vec{E}_0 =
 -\vec{\nabla}\Phi_0+\mathrm j\omega\vec{A}_0,
 \qquad
 \vec{B}_0 =
 \vec{\nabla}\times\vec{A}_0.
 \label{eq:edm-fields-from-potentials}
\end{equation}
One obtains the compact exact result
\begin{equation}
 \calF_0(\vec r)  =
 -\frac{1}{\epsilon_0}\,
 \vec{\sigma}\cdot \Big[ k^2 \, G_k(\vec r) \,\vec{p}_0
 + \vec{\nabla}(\vec{p}_0\cdot\vec{\nabla} G_k(\vec r))
 + \mathrm i \, \mathrm j \, k\, 
 \vec{\nabla} G_k(\vec r) \times\vec{p}_0 \Big] \, .
 \label{eq:edm-field-quaternion-compact}
\end{equation}
The electric and magnetic amplitudes can thus be read off independently:
\begin{subequations}
\begin{align}
 \vec{E}_0 =& \; \frac{1}{\epsilon_0} \left[
 k^2 \; G_k(\vec r) \; \vec{p}_0 + 
 \vec{\nabla}(\vec{p}_0\cdot\vec{\nabla} G_k(\vec r)) \right],
 \label{eq:edm-electric-compact}
\\
 \vec{B}_0(\vec r) =& \; -\mathrm j\mu_0\omega\,
 \vec{\nabla} G_k(\vec r) \times\vec{p}_0 \,.
 \label{eq:edm-magnetic-compact}
\end{align}
\end{subequations}

{\em Electric and Magnetic Dipole Fields.---}For 
a radial function $G_k = G_k(r)$,
one can easily show that
\begin{equation}
 \vec{\nabla}(\vec{p}_0\cdot\vec{\nabla} G_k)
 = \left( G_k''-\frac{G_k'}r \right)
 \vec{n}(\vec{n}\cdot\vec{p}_0)
 + \frac{G_k'}r\,\vec{p}_0.
 \label{eq:edm-hessian}
\end{equation}
Using the relations $G_k' = \left( \mathrm jk-\frac1r \right)G_k$, and
$G_k'' = \left( -k^2-\frac{2\mathrm jk}{r}+\frac2{r^2} \right)G_k$,
one concludes from Eq.~\eqref{eq:edm-electric-compact} that
\begin{align}
\vec{E}_0(\vec{r}) =& \;
 \frac{\ee^{\mathrm jkr}}{4\pi\epsilon_0}
 \Bigg\{ \frac{k^2}{r}
 \left[ \vec{p}_0-\vec{n}(\vec{n}\cdot\vec{p}_0)
 \right] + \left( \frac1{r^3}-\frac{\mathrm jk}{r^2} \right)
 \left[ 3\vec{n}(\vec{n}\cdot\vec{p}_0)-\vec{p}_0 \right] \Bigg\} 
 \qquad (r>0) \,,
 \label{eq:edm-electric-full}
\\
 \vec{B}_0(\vec{r}) =& \;
 \frac{\ee^{\mathrm jkr}}{4\pi\epsilon_0c}
 \left( \frac{k^2}{r}+\frac{\mathrm jk}{r^2}
 \right)
 \vec{n}\times\vec{p}_0
 \qquad (r>0) \, .
 \label{eq:edm-magnetic-full}
\end{align}
Outside the source, the same electric field follows from the
Amp\`ere--Maxwell equation:
\begin{equation}
\vec{\nabla}\times\vec{B}_0
= -\frac{\mathrm j\omega}{c^2}\vec{E}_0,
\qquad
\vec{E}_0 = \frac{\mathrm jc^2}{\omega}
 \vec{\nabla}\times\vec{B}_0.
 \label{eq:edm-electric-from-magnetic}
\end{equation}
Thus calculating \(\vec{E}_0\) directly from the potentials or from
the curl of \(\vec{B}_0\) gives the same complete result.

The ordinary expressions \eqref{eq:edm-electric-full} and
\eqref{eq:edm-magnetic-full} hold for $r>0$. The ideal source requires their
distributional extension at the origin. The relevant identity, with a
spherical principal-value prescription, is
\begin{equation}
 \partial_a\partial_b\frac{1}{4\pi r}
 =\PV\frac{3n_an_b-\delta_{ab}}{4\pi r^3}
 -\frac13\delta_{ab}\delta^{(3)}(\br),
 \label{eq:dipole-hessian-distribution}
\end{equation}
where $\PV$ denotes the principal value and the 
reasoning is described 
in Refs.~\cite{Frahm1983,Hnizdo2011}. Since
$G_k(r)-(4\pi r)^{-1}$ is less singular, the coefficient of its delta
contribution is unchanged. Consequently, the complete electric field is
the spherical distributional extension of
Eq.~\eqref{eq:edm-electric-full}, together with
\begin{equation}
 \vec{E}_{0,\mathrm{contact}}
 =-\frac{\vec{p}_0}{3\epsilon_0}\delta^{(3)}(\br).
 \label{eq:edm-contact}
\end{equation}
This term is \emph{already included} when
Eq.~\eqref{eq:edm-electric-compact} is interpreted distributionally; it
must not be added to that derivative formula a second time.

There is no corresponding delta-function term to append to
Eq.~\eqref{eq:edm-magnetic-compact}. Its kernel contains only the first
derivative of $G_k$, which is locally integrable. The contact term above
belongs to the electric field of an electric point dipole, not to its
magnetic field. At the source, the full Amp\`ere--Maxwell equation includes
$\mu_0\vec{J}_0$; the source-free relation
\eqref{eq:edm-electric-from-magnetic} cannot be used there without that
current contribution. This is also consistent with the distributional
Gauss law $\vec{\nabla}\cdot\vec{E}_0=\rho_0/\epsilon_0$.

{\em Radiated Power.---}The 
physical fields are recovered from their phasors by
Eq.~\eqref{eq:phasor_fields_j}. With peak-amplitude conventions, their
cycle-averaged energy flux is
\begin{equation}
 \langle\vec{S}\rangle=\frac{1}{2\mu_0}
 \operatorname{Re}_{\jj}(\vec{E}_0\times\vec{B}_0^{*_{\jj}}).
 \label{eq:mean-poynting}
\end{equation}
Only phasor conjugation enters this formula; the intrinsic Pauli unit is
not conjugated when extracting physical field amplitudes.

The radiation fields immediately give the power transported to infinity.
For the ideal monochromatic dipole there is, in addition, an exact statement
about its radial mean flux at \emph{every} $r>0$. To verify it using the
full fields, define
\begin{equation}
 \vec{p}_{0,\perp}=\vec{p}_0-\vec{n} ( \vec{n}\cdot\vec{p}_0 ),
 \qquad
 h=\frac1{r^3}-\frac{\jj k}{r^2}.
\end{equation}
Equations~\eqref{eq:edm-electric-full} and
\eqref{eq:edm-magnetic-full} take the form
\begin{align}
 \vec{E}_0&= \frac{1}{4\pi\epsilon_0} \ee^{\jj kr}
 \left[\alpha \, \vec{p}_{0,\perp} +
   2 \; h \; \vec{n} \; ( \vec{n}\cdot\vec{p}_0 ) \right],&
 \alpha&=\frac{k^2}{r}-\frac1{r^3}+\frac{\jj k}{r^2},\\
 \vec{B}_0&= \frac{1}{4\pi\epsilon_0 \, c} \ee^{\jj kr}\,\beta\,
 \vec{n}\times\vec{p}_0,&
 \beta&=\frac{k^2}{r}+\frac{\jj k}{r^2}.
 \label{eq:power-coefficients}
\end{align}
The radial electric term gives no {\em radial} Poynting component. Since
$\alpha\beta^{*_{\jj}}=\frac{k^4}{r^2}+\frac{\jj k}{r^5}$,
the exact complex radial flux is
\begin{equation}
 \frac{\vec{n}\cdot(\vec{E}_0\times\vec{B}_0^{*_{\jj}})}{2\mu_0}
 =\frac{c}{32\pi^2\epsilon_0}
 \left(\frac{k^4}{r^2}+\frac{\jj k}{r^5}\right)
 |\vec{n}\times\vec{p}_0|_{\jj}^2,
 \qquad |\vec{u}|_{\jj}^2=\vec{u}^{*_{\jj}}\cdot\vec{u}.
 \label{eq:complex-radial-flux}
\end{equation}
Taking the $\jj$-real part gives
\begin{equation}
 r^2\vec{n}\cdot\langle\vec{S}\rangle
 =\frac{\dd\langle P\rangle}{\dd\Omega}
 =\frac{\omega^4}{32\pi^2\epsilon_0c^3}
 |\vec{n}\times\vec{p}_0|_{\jj}^2,
 \qquad r>0.
 \label{eq:dipole-angular-power}
\end{equation}
Integration over the solid angle gives the 
total radiated power $\langle P\rangle$ over all directions,
\begin{equation}
 \int\dd\Omega\,|\vec{n}\times\vec{p}_0|_{\jj}^2
 =\frac{8\pi}{3}\vec{p}_0^{*_{\jj}}\cdot\vec{p}_0 \,,
\qquad
 \langle P\rangle
 =\frac{\omega^4}{12\pi\epsilon_0c^3}
 \vec{p}_0^{*_{\jj}}\cdot\vec{p}_0.
 \label{eq:dipole-total-power}
\end{equation}
The factors of $1/2$ are appropriate to peak, not root-mean-square,
phasor amplitudes. The familiar dipole power is therefore recovered while
the exact fields retain their induction and quasistatic parts.

\subsection{Application III: Unified Maxwell Stress Tensor}
\label{sec3C}

After the discussion of the oscillating dipole,
where we met the phasor $\exp(-\jj \omega t)$,
let us return to $\jj$-real fields 
and consider the equation
\begin{equation}
\partial_\mu \mathcal T^\mu 
=-\frac1c\operatorname{Herm}(\mathcal J\mathcal F^\ddagger)  \,,
\qquad \mathcal T^\mu =
\frac{\epsilon_0}{2}\mathcal F^\dagger\bar\sigma^\mu\mathcal F \,,
 \qquad
 \bar\sigma^0=\sigma_0\,,
 \qquad \bar\sigma^a=-\sigma_a \,.
\label{eq:em-bal-master}
\end{equation}
where $(\cdot)^\dagger$ denotes the Hermitian adjoint
(obtained by complex $\ii$-conjugation and transposition),
we use the Minkowski metric $\operatorname{diag}(1,-1,-1,-1)$,
with $x^0=ct$ and $\partial_0=c^{-1}\partial_t$,
and we define the Hermitian part $\operatorname{Herm}(X)$
of a matrix as
\begin{equation}
 \operatorname{Herm}(X)=\frac{X+X^\dagger}{2} \,.
\end{equation}
One easily shows that 
\begin{equation}
 \mathcal T^0=u \, \sigma_0+\frac1c\vec\sigma\cdot\vec S, 
 \qquad 
 u=\frac{\epsilon_0}{2}(E^2+c^2B^2),
 \qquad 
 \mathcal T^a=\frac{S_a}{c}\sigma_0-\sum_bT_{ab} \, \sigma_b \,,
 \label{eq:em-bal-current-components}
\end{equation}
where $a,b\in\{1,2,3\}$ are spatial indices, and
\begin{equation}
 \vec S=\frac1{\mu_0}\vec E\times\vec B,
 \qquad \mathsf T=(T_{ab}),
 \qquad T_{ab}=\epsilon_0(E_aE_b+c^2B_aB_b)-u\delta_{ab}.
 \label{eq:em-bal-Maxwell-stress}
\end{equation}
Here $\mathsf T$ is the ordinary Maxwell stress tensor, with
$(\vec\nabla\cdot\mathsf T)_b=\sum_a\partial_aT_{ab}$.
For each fixed $\mu$, the Hermitian $2\times2$ matrix
$\mathcal T^\mu$ has four real Pauli coefficients:
\begin{equation}
 \mathcal T^\mu=T_{\mathrm{em}}^{\mu0}\sigma_0
            +\sum_{b=1}^3T_{\mathrm{em}}^{\mu b}\sigma_b.
 \label{eq:em-bal-Pauli-coefficients}
\end{equation}
The four choices $\mu=0,1,2,3$ give the four rows of the
$4\times4$ electromagnetic stress--energy tensor.
The subscript ``em'' only identifies the electromagnetic contribution;
this is the same tensor often denoted by $T^{\mu\nu}$:
\begin{equation}
 \bigl(T_{\mathrm{em}}^{\mu\nu}\bigr)=
 \begin{pmatrix}
  u&S_1/c&S_2/c&S_3/c\\
  S_1/c&-T_{11}&-T_{12}&-T_{13}\\
  S_2/c&-T_{21}&-T_{22}&-T_{23}\\
  S_3/c&-T_{31}&-T_{32}&-T_{33}
 \end{pmatrix}.
 \label{eq:em-bal-four-tensor}
\end{equation}

The sign $T_{\mathrm{em}}^{ab}=-T_{ab}$ is essential: the spatial
momentum-flux block is minus the Maxwell traction stress.
We use customary conventions outlined in 
Section~1.3.2 and Section 8.3.2
of Ref.~\cite{Jentschura2017}. For absolute clarity, 
we remark that $T_{ab}$ denotes the separately defined
$3\times3$ Maxwell stress tensor, not an index-lowered
$T_{\mathrm{em}}^{ab}$.
This is a convention that must be kept fixed
when interpreting the force balance.
One finds that 
\begin{equation}
 \mathcal J\mathcal F^\ddagger
 = (\vec J\cdot\vec E-\ii c\vec J\cdot\vec B)\sigma_0
 + \vec\sigma \cdot \left[ c(\rho\vec E+\vec J\times\vec B)
 + \ii \, (\vec J\times\vec E-c^2\rho\vec B)
 \right].
 \label{eq:em-bal-source-expanded}
\end{equation}
All physical field and source components here are real. Taking the
Hermitian part consequently removes precisely the
$\ii$-imaginary Pauli coefficients. Here,
$\operatorname{Re}_{\ii}Q=[Q+C_{\ii}(Q)]/2$ keeps the Pauli basis fixed;
it is not the entrywise real part of the matrix:
\begin{equation}
 \operatorname{Herm}(\mathcal J\mathcal F^\ddagger)
 = \operatorname{Re}_{\ii}(\mathcal J\mathcal F^\ddagger)
 = w\sigma_0+c\vec\sigma\cdot\vec f_{\mathrm L} \,,
\qquad
w = \vec J\cdot\vec E \,,
\qquad
\vec f_{\mathrm L} = \rho\vec E+\vec J\times\vec B \,,
 \label{eq:em-bal-source-Herm}
\end{equation}
where $\vec J\cdot\vec E$ is the power density,
and $\vec f_{\mathrm L} $ is the Lorentz force density.
One inserts Eqs.~\eqref{eq:em-bal-current-components} and
\eqref{eq:em-bal-source-Herm} into Eq.~\eqref{eq:em-bal-master}. 
The result is
\begin{equation}
 \frac1c(\partial_tu+\vec\nabla\cdot\vec S)\sigma_0
  + \vec\sigma\cdot\left[ c^{-2} \partial_t\vec S -
      \vec\nabla\cdot\mathsf T \right]
 =-\frac{w}{c}\sigma_0 -\vec\sigma\cdot\vec f_{\mathrm L}.
 \label{eq:em-bal-components}
\end{equation}
In view of the linear independence of the 
identity matrix and the three Pauli matrices,
the equality of their coefficients gives 
a scalar and a vector equation,
with a minus sign in the vector equation because
$\vec f_{\mathrm L}$ denotes the force density acting on matter:
\begin{equation}
\partial_tu+\vec\nabla\cdot\vec S = - \vec J\cdot\vec E \,,
\qquad
c^{-2} \, \partial_t\vec S - \vec\nabla\cdot\mathsf T=-\vec f_{\mathrm L}.
\label{eq:em-bal-momentum}
\end{equation} 
The scalar equation is Poynting's theorem. The commonly used
Maxwell stress tensor form of Eq.~\eqref{eq:em-bal-momentum} is
\begin{equation}
 \vec\nabla\cdot\mathsf T
 = \rho\vec E+\vec J\times\vec B
   + \frac{\partial}{\partial t}\left(\frac{\vec S}{c^2}\right)
 = \vec f_{\mathrm L}
   + \frac{\partial}{\partial t}\left(\frac{\vec S}{c^2}\right),
\end{equation}
The term proportional to the 
time derivative of the Poynting vector 
is the rate of change of momentum stored in the field.
It is required in a time-dependent problem.
The resulting force law agrees with Eq.~(1.122) of
Ref.~\cite{Jentschura2017}.

\section{Conclusions}
\label{sec4}

The Pauli realization of the complex quaternion algebra provides a direct
retarded inverse of the first-order Maxwell operator. 
The quaternionic formalism simplifies and unifies
a number of calculations in electromagnetic theory. 
Its construction
rests on the scalar factorization $\calD\calD^\ddagger=\Boxop\sigma_0$.
In momentum space [see Eq.~\eqref{eq:Pinverse}],
this gives a result for the 
quaternionic Maxwell Green function whose structure
resembles the Feynman propagator for fermion fields;
the retarded prescription fixes the boundary condition at the
light cone. Charge conservation removes the unwanted
scalar Pauli component of the quaternionic field $\calF$.

There are two equivalent routes through the calculation. The direct route
is $\calF=(\epsilon_0c)^{-1}\calG_R*\calJ$. The potential route is
$\calA_R=\mu_0G_R*\calJ$, followed by $\calF=c\calD^\ddagger\calA_R$.
The second route naturally satisfies Lorenz gauge.
The first route requires no gauge choice for the
fields themselves. In either route, a retarded boundary condition and
appropriate source behavior are part of the solution specification.

The applications test different aspects of this structure. 
For a moving charge, the retarded-time
constraint supplies the Li\'enard--Wiechert Jacobian. For the oscillating
dipole, retaining the complete Helmholtz kernel before differentiation
recovers the near, induction, and radiation terms and verifies Lorenz
gauge exactly. Its mean radial power agrees with the radiation-zone
result at every radius outside the ideal source, although its full local
energy flux contains near-field structure.
As a further application, we present a streamlined formulation
of the Maxwell stress tensor, naturally combined 
with Poynting's theorem in view of the quaternionic structure. 

The additional central unit $\jj$ separates Fourier and phasor phase from
the intrinsic $\ii$ of the Pauli algebra. This gives a faithful harmonic
extension of the single-field representation; it is a choice of algebraic
organization, not a modification of Maxwell theory. Appendix~\ref{app:quaternions}
fixes the noncommutative Hamilton conventions, while
Appendix~\ref{app:bicomplex} makes the independent commuting complex
structures explicit as real matrices.

The inverse-operator approach is related to earlier quaternionic causal
fundamental solutions~\cite{Grudsky2004}, time-harmonic 
constructions~\cite{DelgadoKravchenko2023}, and the recent Fourier-based biquaternionic
preprint \cite{GuillenVillalobos2026}. The present treatment emphasizes the
connection between Edmonds's compact equation, explicit SI-normalized
retarded kernels, conservation and gauge identities, and fully worked
static and radiative examples. 
These complementary perspectives also
connect the electromagnetic analysis with the matrix representations
and commutative quaternion structures studied in
Refs.~\cite{KatrusiakLe2024,Chen2025,Chang2026}.

\section*{Acknowledgments}

This work was supported by the National Science Foundation through
Grant PHY-2513220 (U.D.J.).
The authors acknowledge extremely helpful and insightful 
conversation with {\tt gpt-5.6-high} in regard to the 
bicomplex algebra for oscillating fields,
and assistance in checking many of the calculations reported here.
The authors independently performed
and verified all calculations, interpretations, and scientific
conclusions presented in this paper,
and assume full and complete responsibility for the work.

\section*{Author Contributions}

I.H.: Verification of all equations in the paper;
contributions to the writing of the paper. 
Conceptualization of the extension of 
the quaternionic formalism to the Maxwell stress tensor.
B.S.: Partial verification of the results 
communicated and help in the clarification of the literature background.
U.D.J.: Conceptualization, formulation of basic principles,
calculation of the Green function and conceptualization
of the example applications, writing the paper.

\section*{Data Availability}

The original contributions presented in this study are included in the 
article. Further inquiries can be directed to the corresponding author.

\appendix

\section{Aspects of Quaternions}
\label{app:quaternions}

\emph{Complex Numbers and Matrices.---}
The real representation of an ordinary complex number is not unique.
For consistency with Appendix~\ref{app:bicomplex}, we adopt
\begin{equation}
 K=\begin{pmatrix}0&1\\-1&0\end{pmatrix}=\ii\sigma_2,
 \qquad K^2=-I_2,
 \qquad
 \MM(x+\ii y)=xI_2+yK=\begin{pmatrix}x&y\\-y&x\end{pmatrix}.
 \label{eq:complex-real-rep-A}
\end{equation}
It follows immediately that
\begin{equation}
 \MM(z_1z_2)=\MM(z_1)\MM(z_2).
\end{equation}
For a unit-modulus complex number,
\begin{equation}
 \MM(\e^{\ii\theta})=
 \begin{pmatrix}\cos\theta&\sin\theta\\-\sin\theta&\cos\theta\end{pmatrix},
\end{equation}
which is a rotation through $-\theta$ when acting actively on column
vectors in the usual oriented plane. In this convention $\ii$ is a
$-90^\circ$ rotation and its square a $180^\circ$ rotation, $-I_2$.
The equally valid convention with both off-diagonal signs reversed is
related by the real change of basis $\operatorname{diag}(1,-1)$.
Thus the sign of the rotation assigned to the imaginary unit is a
representation convention, not an algebraic difference.

\emph{Quaternions and Matrices.---}
The Pauli matrices are
\begin{equation}
 \sigma_0=\begin{pmatrix}1&0\\0&1\end{pmatrix},\quad
 \sigma_1=\begin{pmatrix}0&1\\1&0\end{pmatrix},\quad
 \sigma_2=\begin{pmatrix}0&-\ii\\\ii&0\end{pmatrix},\quad
 \sigma_3=\begin{pmatrix}1&0\\0&-1\end{pmatrix}.
\end{equation}
Hamilton's right-handed quaternion units may be chosen as
\begin{equation}
 e_a=-\ii\sigma_a,\qquad
 e_ae_b=-\delta_{ab}\sigma_0+\varepsilon_{abc}e_c.
 \label{eq:hamilton-units}
\end{equation}
They anticommute for distinct indices, unlike the independent commuting
units of Appendix~\ref{app:bicomplex}. Real quaternions are represented by
\begin{equation}
 q=q_0\sigma_0-\ii\vec{q}\cdot\bsig,\qquad
 r=r_0\sigma_0-\ii\vec{r}\cdot\bsig,
 \qquad q_0,q_a,r_0,r_a\in\RR.
\end{equation}
The Pauli identity
\begin{equation}
 (\bsig\cdot\vec{a})(\bsig\cdot\vec{b})
 =(\vec{a}\cdot\vec{b})\sigma_0
 +\ii\bsig\cdot(\vec{a}\times\vec{b})
 \label{eq:PauliIdentity}
\end{equation}
gives
\begin{equation}
 qr=(q_0r_0-\vec{q}\cdot\vec{r})\sigma_0
 -\ii\bsig\cdot(q_0\vec{r}+r_0\vec{q}+\vec{q}\times\vec{r}).
 \label{eq:hamilton-product}
\end{equation}
Equivalently, in scalar--vector pair notation,
\begin{equation}
 (q_0,\vec{q})(r_0,\vec{r})
 =(q_0r_0-\vec{q}\cdot\vec{r},\,
 q_0\vec{r}+r_0\vec{q}+\vec{q}\times\vec{r}).
\end{equation}
This displays the scalar and cross products responsible for the
quaternionic form of the divergence and curl operations.
Quaternionic conjugation reverses all three Hamilton units:
\begin{equation}
 q^\ddagger=q_0\sigma_0+\ii\vec{q}\cdot\bsig,
 \qquad qq^\ddagger=(q_0^2+q_1^2+q_2^2+q_3^2)\sigma_0.
\end{equation}
For a nonzero real quaternion,
\begin{equation}
 |q|^2=q_0^2+q_1^2+q_2^2+q_3^2,
 \qquad q^{-1}=\frac{q^\ddagger}{|q|^2}.
 \label{eq:quaternion-inverse}
\end{equation}
After complexification, the coefficients are allowed to be complex and
the algebra is isomorphic to $M_2(\CC_{\ii})$. A useful basis is then
$\sigma_0,\sigma_1,\sigma_2,\sigma_3$, as in the main text. For
$Q=q_0\sigma_0+\bsig\cdot\vec{q}$ one has
$QQ^\ddagger=(q_0^2-\vec{q}^{\,2})\sigma_0$; this quadratic form can vanish
for a nonzero complex quaternion. Thus the real-quaternion division
property must not be transferred to the complex algebra. The light-cone
singularity in the Maxwell inverse is an instance of this distinction.
For further matrix and transformation applications, see
Refs.~\cite{Edmonds1978,KatrusiakLe2024}.

{\em Further Generalizations.---}The standard real 
normed division algebras are the reals $\RR$, 
the complex numbers $\CC$, the
quaternions $\mathbb H$, and the octonions $\mathbb O$, of dimensions
$1$, $2$, $4$, and $8$, respectively (see also Table~\ref{table1}). The classification by a
multiplicative positive norm is the Hurwitz theorem; related topological
restrictions on finite-dimensional real division algebras were obtained
by Bott and Milnor \cite{BottMilnor1958}. These statements should not be
confused with a uniqueness classification of all possible nonassociative
real division algebras. The associative finite-dimensional real division
algebras are only $\RR$, $\CC$, and $\mathbb H$. A detailed account of
these distinctions and of octonions is given by Baez \cite{Baez2002}.

Octonion multiplication is nonassociative, so there is no faithful
multiplication-preserving embedding of the octonion algebra into an
ordinary associative matrix algebra. Nevertheless, left multiplication
by an octonion is an $8\times8$ real linear operator. Those operators do
not in general satisfy $L_aL_b=L_{ab}$, and hence are not an associative
matrix realization of the octonion product. The issue is the preservation
of multiplication, not the possibility of representing individual linear
maps by matrices.

\begin{table}[tb]
\caption{Properties of the four real normed 
division algebras
(real numbers $\mathbb R$, 
complex numbers $\mathbb C$, 
quaternions $\mathbb H$, and 
octonions $\mathbb O$;
see Ref.~\cite{Baez2002}, especially Secs.~1.1 and 2.2).
The restriction to dimensions $1$, $2$, $4$, and $8$ holds
more generally for nonzero finite-dimensional real algebras
with bilinear multiplication and no zero divisors
\cite{BottMilnor1958}.
``Ordered'' means a total order compatible with addition
and multiplication and extending the usual order on
$\mathbb R$.
The table concerns the listed normed division algebras,
not every real division algebra of the corresponding
dimension.}
\label{table1}
\begin{ruledtabular}
\begin{tabular}{lccccc}
Algebra & Dimension & No Zero Divisors & Associative & Commutative & Ordered\\
$\RR$       &1&Yes&Yes&Yes&Yes\\
$\CC$       &2&Yes&Yes&Yes&No\\
$\mathbb H$ &4&Yes&Yes&No&No\\
$\mathbb O$ &8&Yes&No&No&No
\end{tabular}
\end{ruledtabular}
\end{table}

\section{Aspects of Bicomplex Numbers}
\label{app:bicomplex}

{\em Bicomplex Imaginary Units.---}For 
the coefficient algebra used for harmonic fields in
Sec.~\ref{sec3}, we introduce two independent commuting imaginary units
\begin{equation}
 \ii^2=-1,
 \qquad
 \jj^2=-1,
 \qquad
 \ii \, \jj=\jj \, \ii,
 \label{eq:bicomplex_rules}
\end{equation}
and we are now in search of a matrix representation
of this algebra.
The resulting bicomplex algebra is
\begin{equation}
 \BB = \RR[\ii,\jj] = \left\{ z = a+b\ii+c\jj+d\,\ii\jj
 \; ;\; a,b,c,d\in\RR
 \right\}.
 \label{eq:Bdef}
\end{equation}
Equivalently,
 $\BB \simeq \CC_{\ii}\otimes_{\RR}\CC_{\jj}$
can be written in terms of a product of two 
complex algebras under real multiplication.
A general bicomplex number may be grouped according to the
\(\ii\)-complex structure,
\begin{equation}
 z = {\rm Re}_{\ii}(z) + \ii \; {\rm Im}_{\rm i}(z) \,, 
 \qquad
 {\rm Re}_{\ii}(z) = a+c \, \jj,
 \qquad
 {\rm Im}_{\rm i}(z) = b+d \, \jj \,.
\label{eq:z_outer_i}
\end{equation}
This observation suggests applying the ordinary complex matrix
representation once again:
\begin{equation}
 z=z_R+\ii z_I
 \quad\longleftrightarrow\quad
 \begin{pmatrix}
 z_R&z_I\\
 -z_I&z_R
 \end{pmatrix},
 \label{eq:complex_block}
\end{equation}
where the entries of this $2\times2$ matrix are now
$\jj$-complex numbers.
This bicomplex representation naturally leads to 
$4 \times 4$-matrices.

{\em Examples of Matrix Representations.---}According 
to Eq.~\eqref{eq:complex_block},
the explicit form of a matrix representation
of a bicomplex number is
\begin{equation}
 \MM_{\BB}
 \left(
 a+b\ii+c\jj+d\,\ii\jj
 \right)
 =
 \begin{pmatrix}
 a&c&b&d\\
 -c&a&-d&b\\
 -b&-d&a&c\\
 d&-b&-c&a
 \end{pmatrix}.
 \label{eq:full_rep}
\end{equation}
The representation is faithful and multiplicative,
$\MM_{\BB}(z_1z_2) = \MM_{\BB}(z_1)\MM_{\BB}(z_2)$.
The tensor-product form makes the two complex structures especially
transparent.  One finds that
\begin{equation}
 \MM_{\BB}(\ii) =
 \begin{pmatrix}
 0&0&1&0\\
 0&0&0&1\\
 -1&0&0&0\\
 0&-1&0&0
 \end{pmatrix},
 \label{eq:i_matrix}
\end{equation}
while
\begin{equation}
 \MM_{\BB}(\jj) =
 \begin{pmatrix}
 0&1&0&0\\
 -1&0&0&0\\
 0&0&0&1\\
 0&0&-1&0
 \end{pmatrix}.
 \label{eq:j_matrix}
\end{equation}
Straightforward calculations lead to 
$\MM_{\BB}(\ii)^2 = \MM_{\BB}(\jj)^2 = -\mathbbm{1}_{4 \times 4}$,
and the two matrix representations commute:
$[\MM_{\BB}(\ii),\MM_{\BB}(\jj)]=0$.
The product is represented by
\begin{equation}
 \MM_{\BB}(\ii\, \jj) =
 \begin{pmatrix}
 0&0&0&1\\
 0&0&-1&0\\
 0&-1&0&0\\
 1&0&0&0
 \end{pmatrix}.
 \label{eq:ij_matrix}
\end{equation}
Since $(\ii\jj)^2 = \ii^2\jj^2 = +1$,
we correspondingly have
$\MM_{\BB}(\ii\jj)^2= \mathbbm{1}_{4 \times 4}$.

{\em Bicomplex Inverse.---}The 
bicomplex algebra contains the two idempotents
\begin{equation}
 e_\pm = \frac{1}{2} \left( 1\pm\ii\jj \right).
 \label{eq:idempotents}
\end{equation}
They satisfy $e_\pm^2=e_\pm$, $e_+e_-=0$, and $e_++e_-=1$.
One can decompose a bicomplex number according to 
\begin{subequations}
\begin{align}
 z=& \; z_+e_++z_-e_-, \\
 z_+ =& \; (a+d)+\ii(b-c), \\
 z_- =& \; (a-d)+\ii(b+c).
 \label{eq:zpm}
\end{align}
\end{subequations}
If $z_+\ne0$ and $z_-\ne0$, the inverse $z^{-1}$ is then easily obtained as
\begin{equation}
 z^{-1} = z_+^{-1}e_+ + z_-^{-1}e_-,
 \label{eq:bicomplex_inverse}
\end{equation}
and correspondingly $\MM_{\BB}(z^{-1}) = \MM_{\BB}(z)^{-1}$
in the faithful matrix representation. 
In summary, we notice
that applying the real $2\times2$ representation of complex numbers twice
produces a faithful $4\times4$ representation of two commuting complex
structures. Related commutative quaternion matrix problems
are treated in Refs.~\cite{Chen2025,Chang2026}.


%

\end{document}